\documentclass[conference]{IEEEtran}
\IEEEoverridecommandlockouts

\usepackage{cite}
\usepackage{multicol}

\usepackage{amsmath}
\usepackage[T1]{fontenc}
\usepackage{algpseudocode}
\usepackage[utf8]{inputenc}
\usepackage[english]{babel}
\usepackage[usenames, dvipsnames]{color}
\usepackage{pifont}
\usepackage{booktabs}
\usepackage{amsmath}
\usepackage{mathtools}
\usepackage{textcomp}
 \usepackage{graphicx}
\usepackage{subfigure}
\usepackage[ruled,vlined,commentsnumbered]{algorithm2e}
\usepackage{amsthm}
\usepackage{setspace}

\usepackage{fancyhdr}
\fancypagestyle{firstpage}{%
  \lhead{This work has been accepted for publication in the IEEE WCNC 2024 Conference.}
}

\SetKwInput{KwIn}{Inputs}

\theoremstyle{definition}

\theoremstyle{remark}

\theoremstyle{plain}

\newcommand{\RNum}[1]{\uppercase\expandafter{\romannumeral #1\relax}}
\newcommand{\forallK}{\forall\,k \in \mathcal{K}}

\def\BibTeX{{\rm B\kern-.05em{\sc i\kern-.025em b}\kern-.08em
    T\kern-.1667em\lower.7ex\hbox{E}\kern-.125em}}
    
\begin{document}

\title{Cooperative Multi-Monostatic Sensing for Object Localization in 6G Networks}
\author{\IEEEauthorblockN{ Maximiliano Rivera Figueroa, Pradyumna Kumar Bishoyi, and Marina Petrova}
\IEEEauthorblockA{Mobile Communications and Computing, RWTH Aachen University, Aachen, Germany \\
Email: \{maximiliano.rivera, pradyumna.bishoyi, petrova\}@mcc.rwth-aachen.de}
}

\maketitle
\thispagestyle{firstpage}
\begin{abstract}
Enabling passive sensing of the environment using cellular base stations (BSs) will be one of the disruptive features of the sixth-generation (6G) networks. However, accurate localization and positioning of objects are challenging to achieve as multipath significantly degrades the reflected echoes. Existing localization techniques perform well under the assumption of large bandwidth available but perform poorly in bandwidth-limited scenarios. To alleviate this problem, in this work, we introduce a 5G New Radio (NR)-based cooperative multi-monostatic sensing framework for passive target localization that operates in the Frequency Range 1 (FR1) band. We propose a novel fusion-based estimation process that can mitigate the effect of multipath by assigning appropriate weight to the range estimation of each BS. Extensive simulation results using ray-tracing demonstrate the efficacy of the proposed multi-sensing framework in bandwidth-limited scenarios.
\end{abstract}


\section{Introduction} \label{sec:Introduction}
The upcoming sixth generation (6G) wireless networks is expected to facilitate immersive applications, such as autonomous driving, extended reality (XR), and digital twins, which require the highest data rates and minimal latency along with centimeter-level positioning accuracy and context information about the surrounding objects and the environment \cite{Zhang2021-PMN}. Towards this vision, the third generation partnership project (3GPP) proposed cellular network-based positioning solutions in Rel-16 \cite{Rel16} and has continued to enhance the techniques for 5G new radio (NR) in Rel-17 and 18 \cite{Rel17, Rel18}. The current 5G NR-based positioning system is envisioned in both Frequency Range 1 (FR1), i.e., sub-$6$ GHz, and FR2 in mmWave ($24$-$71$ GHz), and the positioning techniques are based on received downlink and uplink reference signal strength (RSS) and time and angle-based measurements such as time-of-arrival (ToA), time-difference-of-arrival (TDOA), round trip time (RTT), and angle-of-arrival (AoA). Importantly, the positioning system is based on active sensing, in which the target (in this instance, user equipment) is registered to the network and has communication capability to receive and transmit position referencing signal. Future 6G networks, however, will also be required to facilitate precise positioning of passive objects (with no communication interface) to enable location-based services to immersive applications. \textit{The primary focus of our work is to study, propose, and evaluate solutions that ensure precise positioning in the case of passive sensing, which remains an open research problem.}
\begin{figure}[t]
    \centering
    \includegraphics[width=\linewidth, trim={0cm 1.25cm 0cm 0cm},clip]{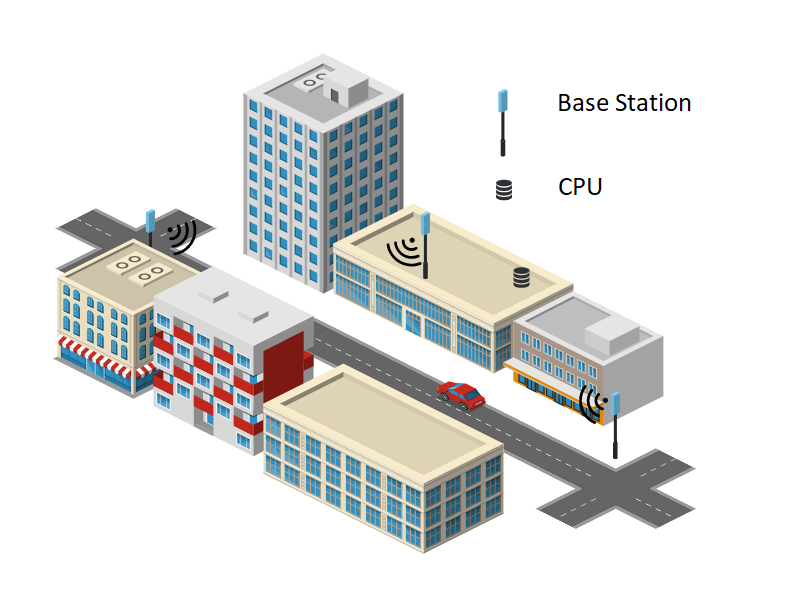}
    \caption{Illustration of an urban environment involving multiple Base Stations (BSs), a target (vehicle), and a Central Processing Unit (CPU).}
    \label{fig:illustration_SystemModel}
    \vspace{-0.5cm}
\end{figure}

One of the fundamental challenges in passive sensing is the erroneous position estimation due to the superposition of multipath components. In passive sensing, the cellular base station (BS) acts as a monostatic radar and relies on the reflected echo signals from the target for localization. In the case of harsh environments, obtaining reliable measurements is challenging due to the non-line-of-sight (NLOS) propagation of reflected echo signals. Several works suggest improving the location estimation by mitigating the effect of NLOS parameters. A comprehensive study of different techniques for mitigation of the NLOS effects is presented in \cite{Guvenc2009}. In \cite{Schieler2020}, an iterative approach is implemented to estimate the contribution of all paths, starting from a single path, for an orthogonal frequency division multiplexing (OFDM) radar receiver. Furthermore, in \cite{Perez-Cruz2016}, an iterative blind learning algorithm is implemented for estimating the error bias due to NLOS reflections. All of the aforementioned techniques implicitly assume that the total number of NLOS paths is known in advance and can be eliminated from the LOS path. \textit{However, in a real scenario, the number of NLOS paths is not known a priori and adds a relatively large NLOS bias in the order of tens of meters and up to hundreds of meters in extreme scenarios, thus significantly reducing the precision of position estimation \cite{venkatraman2004}.}  

A promising alternative to mitigate the multipath effects in the estimation is to leverage the dense deployment of 5G BSs and employ cooperative monostatic sensing, which we call \textit{multi-monostatic sensing}, to localize a passive target accurately. With this approach, multiple BSs independently localize the target and then combine their information to improve the position accuracy and robustness. Moreover, in a dense network scenario, when localizing the target, multiple BSs may have stronger LOS components which help in minimizing the location error due to NLOS components. 
In this work, we analyze the performance of multi-monostatic sensing while taking the effect of multipath propagation into account. Only few works have so far considered fusion for localization in the context of 5G sensing \cite{Mike2017}\cite{Henninger2022}.
However, in contrast to our work, these works rely on active sensing by multiple BSs.  Different from these, in \cite{Shi2022}, authors studied cellular network-based passive sensing, where multiple BSs aim to locate different targets and eliminate ghost readings. The above analysis assumes a $100$ MHz channel bandwidth in the FR1 band for positioning, which may not be available in a practical scenario where the base station must share the channel for communication. In a bandwidth-limited scenario, the precision of the algorithm may degrade. Therefore, in this paper, we address the following queries regarding multi-monostatic sensing: (i) how to effectively fuse different BSs' estimations to minimize the overall positioning error, and (ii) what is the impact of the bandwidth and the number of BSs on the position estimation outcome.

\subsection*{Contributions}
In this work, we introduce a 5G NR-based \textit{multi-monostatic sensing for passive target localization} that operates in the FR1 band. In particular, we consider a two-stage positioning method for localizing a moving target in a multipath-rich environment. In the first stage, each BS, as a monostatic radar, transmits an OFDM signal and captures its echo to estimate the relative distance to a target. We use the ToA to estimate the distance between BS and the target. In the second stage, inspired by \cite{Henninger2022, Shi2022}, the estimates are processed in a central processing unit (CPU) to improve the estimation accuracy due to multipath-induced errors and to find the target's position. For our fusion-based estimation, we consider three different algorithms, namely maximum likelihood (ML), maximum a posteriori (MAP), and non-linear least squares (NLLS) and analyze the impact of fusion on the estimation accuracy. The main contributions of our work are:
\begin{itemize}
    \item We study multi-monostatic sensing for passive target localization and analyze the impact of multipath propagation on the localization. 
    \item We characterize the outcome of the fusion-based estimation process under three different algorithms to limit the effect of multipath. Furthermore, we propose a novel approach for determining the weights for the fusion algorithm based on the outcomes of the periodogram output of each BS. This simplifies the implementation of the fusion algorithm without the use of an iterative technique. 
    \item Finally, we evaluate the efficacy of the fusion algorithms in scenarios with varying bandwidths and numbers of BSs. Our analysis indicates that multi-monostatic sensing improves the range estimation accuracy even in bandwidth-limited scenarios, and the performance of ML and MAP-based estimation is similar for higher bandwidth cases.
\end{itemize}
The rest of the paper is organized as follows. We describe the system model in Section \ref{sec:SystemModel}. In Section \ref{sec:MultiMonostatic}, we describe the framework for multi-monostatic sensing and analyze different fusion algorithms. Further, we show the numerical evaluation in Section \ref{sec:PerformanceEvaluation} and conclude the paper in Section \ref{sec:Conclusions}.
\section{System Model}\label{sec:SystemModel}
We consider a system consisting of $K$ 5G NR BSs located in an urban area, where each of the BSs acts as a monostatic radar with a full-duplex sensing capability \cite{Zhang2021-PMN}, operating in the FR1 band. 
We assume that each BS uses orthogonal resources for sensing, i.e., there is no inter-BS interference. Further, all the BSs are synchronized and connected to a CPU via a high-capacity backhaul link. Fig.~\ref{fig:illustration_SystemModel} depicts an illustration of the system model.
The $K$ BSs aim to locate a single target of unknown 3D position, denoted by $\mathbf{x} = [x,\,y,\,z]^T$, moving with an unknown constant speed of $\mathbf{v}$ m/s. Each BS is at a known position $\mathbf{x}_k = [x_k,\, y_k, z_k]^T$, with $k \in \mathcal{K}$, and $\mathcal{K} = \{1, \dots, K\}$. We use the terms target and object interchangeably in our work.

Each BS independently estimates its relative distance $\hat d_k$ to the target by sending an OFDM pilot signal, processing the received echo, and computing the ToA. The present analysis focuses on a single target; however, it is worth noting that an extension of this study to multiple targets can be easily performed by incorporating an intermediate clustering step, as shown in \cite{Molisch2014}.

\subsubsection{Transmit Signal Model} The transmit signal consists of $N_s$ active subcarriers with $\Delta f$ of subcarrier spacing (SCS) and $M$ OFDM symbols. The baseband time-domain signal is represented as follows \cite{Barneto2019}:
\begin{equation}
s(t) = \sum_{m=0}^{M-1} \sum_{n=0}^{N_s-1} c_{m,\,n}\, e^{j\,2\pi n \Delta f\,t} g(t - mT)
\label{eq:OFDM_Tx_Signal}
\end{equation}
where $c_{m,\,n}$ represent the complex modulation symbols for a given symbol $m$ and subcarrier $n$, and $g(\cdot)$ is the pulse shaper. The overall OFDM symbol duration consists of $T = T_{cp} + T_s$, where $T_{cp}$ is the cycle prefix (CP) duration and $T_s$ is the symbol duration. 

\subsubsection{Received Echo Signal} In order to process the received signal, several steps are considered. First, CP removal is performed, followed by the Fourier Transform over each received symbol and applying Zero-Forcing (ZF) to remove the modulation symbols. Then, the received signal of the $m$-th OFDM symbol and $n$-th subcarrier at the $k$-th BS $D_{m, n}^k$, is given by \cite{Barneto2019}:
\begin{equation}
D_{m, n}^k = \gamma^k \,e^{j2\pi\,(m\,T_s f_D^k - n\tau^k \Delta f) }  + N_{m, n}^k,\,\,\forallK \label{eq:OFDM_Rx_ZF}
\end{equation}
where $\gamma^k$ is the radar channel gain which covers the path loss and the radar cross section (RCS) for the $k$-th BS, $f_D^k$ is the Doppler shift seen by the $k$-th BS, and $\tau^k$ is the round-trip propagation delay of the $k$-th BS. $N_{m, n}^k$ is the noise of the $m$-th OFDM symbol and $n$-th subcarrier at the $k$-th BS after ZF.
Furthermore, if multipath is considered, the received signal can be represented as follows
\begin{equation}
    \tilde{D}_{m, n}^k = \sum_{\ell = 0}^L \gamma_{\ell} \,e^{j2\pi\,(m\,T_s f_{D, \ell}^k - n\tau_{\ell}^k \Delta f) } + N_{m, n}^k,\,\,\forallK \label{eq:Multipath_OFDM_Rx}
\end{equation}
where the sub-index $\ell$ indicates the different paths. The term $f_{D,\ell}^k$ represents the Doppler shift of the target of the $\ell$-th path at the $k$-th BS, where it may differ from the line-of-sight (LOS) path due to the movement of other scatterers. 

\subsubsection{ToA and Doppler Shift Estimation} In order to estimate the time-of-arrival (ToA) and the Doppler shift, a periodogram-based approach is implemented at each BS. 
This well-known technique can be directly applied to an OFDM signal by just performing a 2D-FFT over $\tilde{D}_{m, n}^k$, making it a simple signal processing algorithm \cite{Braun2014_1000038892}. Furthermore, the ToA estimation can be mapped to range as each BS acts as a monostatic radar.

Following \cite{Pucci2022}, the periodogram approach used at BS $k \in \mathcal{K}$ to determine range and velocity of a target is expressed as
\begin{multline}
    A_k(n, m)= \Bigg| \sum_{r=0}^{N'_s-1} \sum_{l=0}^{M'-1} \Big(\tilde{D}_{r, l}^k e^{-j2\pi \,l\,m/M'}\Big) e^{j2\pi\,r\,n/N'}\Bigg|^2
    \label{eq:Periodogram}
\end{multline}
where $N'_s \ge N_s$ and $M'\ge M$. Then, the range and velocity estimated at each BS can be computed as follows 
\begin{equation}
    \hat d_k = \frac{\hat n_k c_0}{2 \,\Delta f\, N'_s}, \quad \hat v_k = \frac{\hat m_k c_0}{2\,f_C\, T\, M'},\,\,\forallK \label{eq:Distance_Velocity_Estimation_Periodogram}
\end{equation}
where $\hat n_k$ and $\hat m_k$ are the peaks in the periodogram, which are determined from the following equation \cite{Pucci2022}, 
\begin{equation}
    (\hat n_k,\,\hat m_k) = \text{arg}\max_{n, \,m} A_k(n, m),\,\,\forallK
    \label{eq:PeakSelection}
\end{equation}
Eq. \eqref{eq:PeakSelection} refers to a peak selection method that fulfils the hypothesis test where the signal level is higher than a predefined threshold. Note that in our analysis, we assume that the signal transmission, echo reception, and estimation procedure all take place inside a coherent processing interval (CPI), during which all sensing parameters remain unchanged. Additionally, we assume a clutter-free environment for our analysis. Nevertheless, it is possible to extend our analysis to more complex environments with clutters, by implementing background subtraction techniques \cite{Zhang2021-PMN,tivive2015} to mitigate the effect of clutter.

\begin{figure}[h!]
    \centering
    \includegraphics[angle = -90, width=0.9\linewidth, trim={3cm 2cm 3cm 4cm},clip]{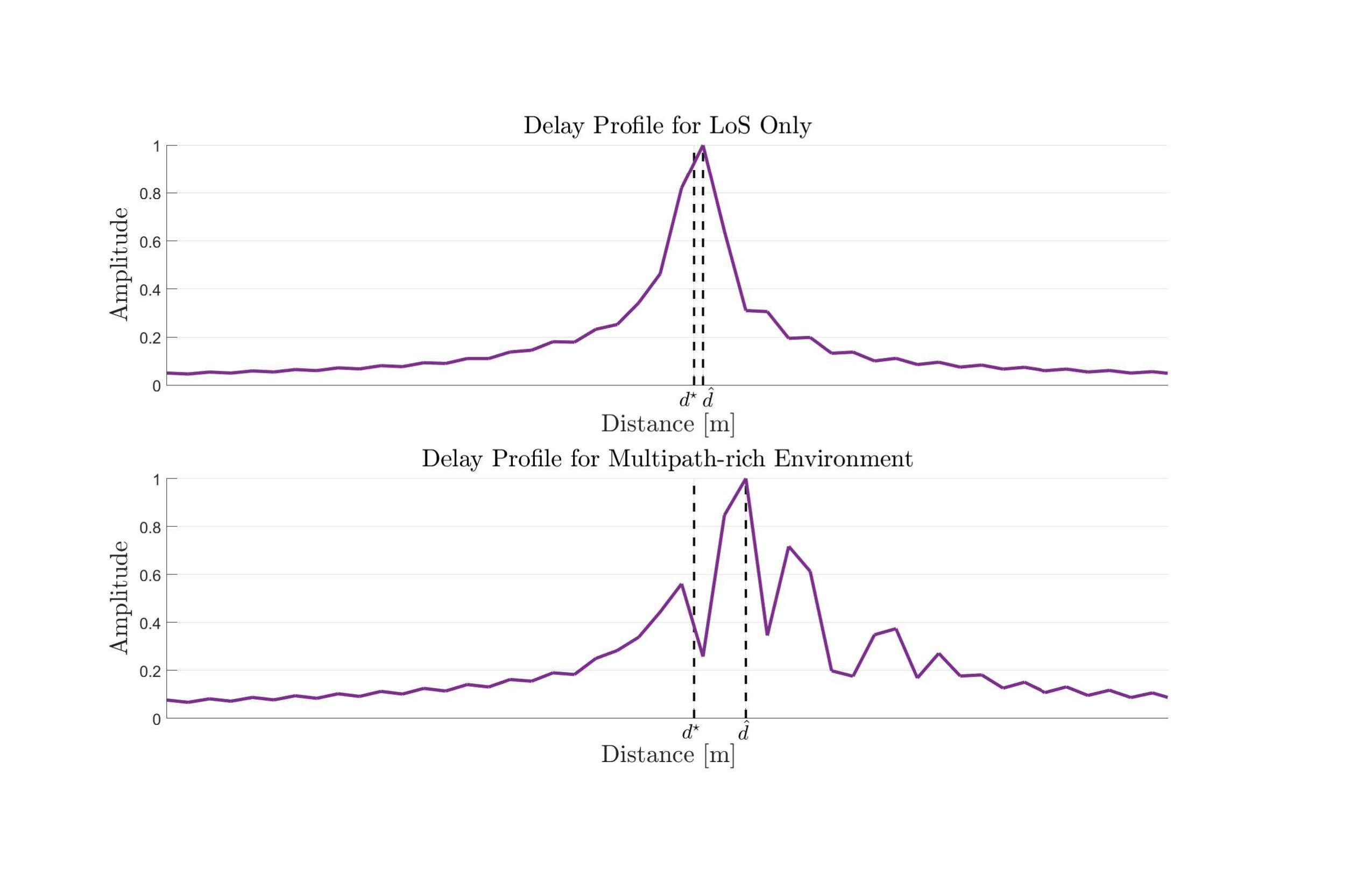}
    \caption{Delay profile comparison between a LOS-only and a multipath-rich environment.}
    \label{fig:Multipath_v_Los_DelayProfile}
\end{figure}

In \cite{Ge2023, Guvenc2009}, it has been shown that the TOA-based localization technique performs well for LOS-only conditions, but its accuracy degrades significantly in multipath-rich environments. Further, from the Eq. \eqref{eq:Multipath_OFDM_Rx}, multipath can be seen as a superposition of different NLOS paths received at the BS at different delays in the periodogram map $A(n,\, m)$, this effect is illustrated in Fig.~\ref{fig:Multipath_v_Los_DelayProfile}. This figure depicts the delay profile for LOS and multipath conditions, where $d^{\star}$ is the true position and $\hat d$ is the estimated distance based on Eq. \eqref{eq:PeakSelection}. It can be seen that the distance error $|\hat d - d^{\star}|$ is higher in a multipath-rich environment. The accuracy of ToA-based localization is limited by the bandwidth of the transmitted signal \cite{Zhang2021-PMN}. This motivates us to opt for collaborative target localization by multiple BSs, as described in detail in the subsequent section. 

\section{Multi-Monostatic Sensing}\label{sec:MultiMonostatic}

Multi-monostatic sensing arises as a solution to decrease the accuracy error by combining the estimates of different BSs that perform monostatic sensing. It works in two different stages. In the first stage, all the BSs estimate their relative distance to the target, while in the second stage, the estimates are fused in a specific manner to improve the accuracy.


Our research delves into various fusion algorithms that aim to enhance accuracy by mitigating multipath-induced estimation errors. To achieve this, the distance estimate from each base station serves as an input for a fusion algorithm that runs on a CPU. The resulting output is the target's position \cite{Henninger2022}. The fusion algorithms considered are described below.
\begin{figure}[t]
    \centering
    \includegraphics[width=0.70\linewidth]{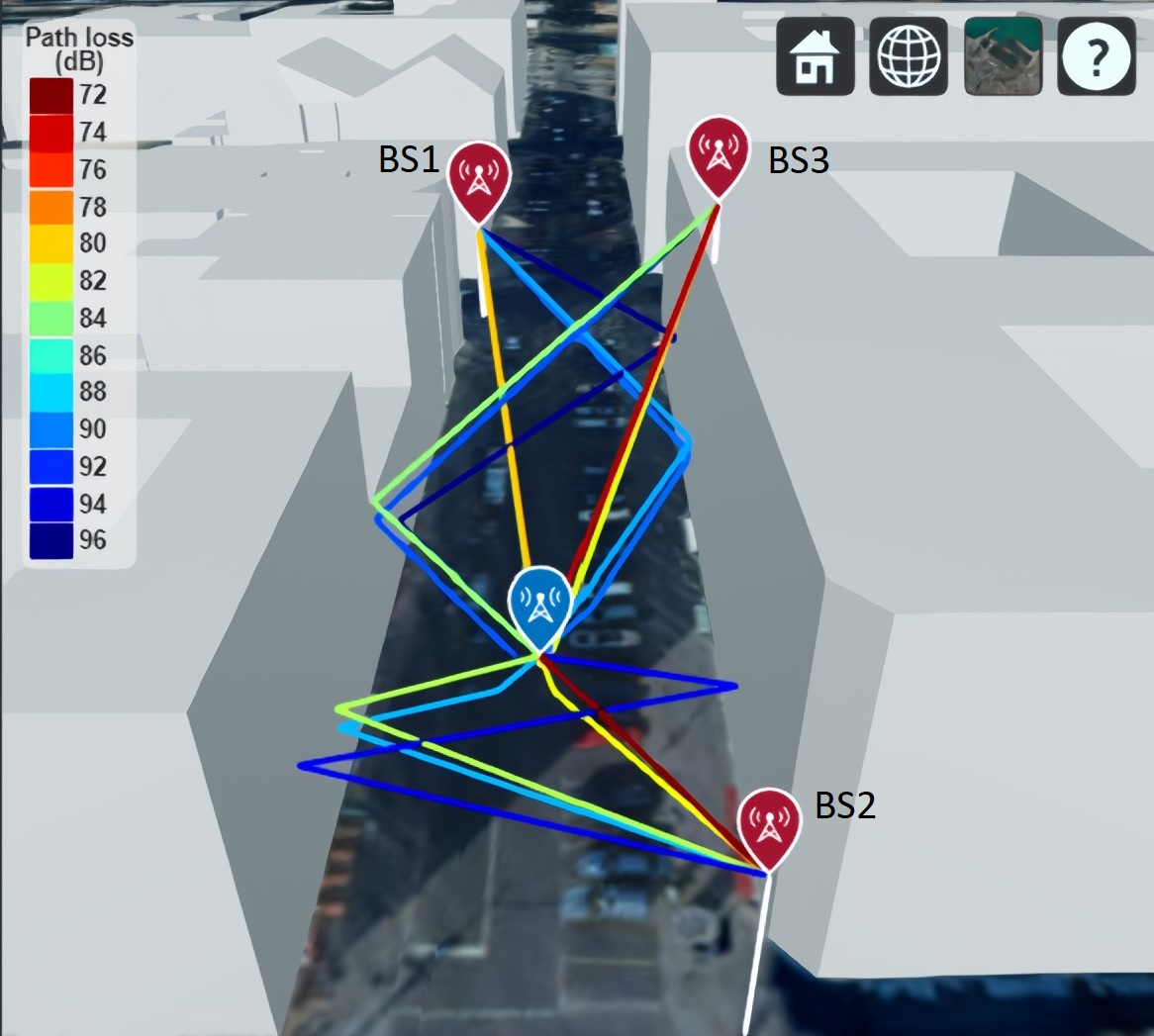}
    \caption{Example of the ray-tracing simulator for a given target's position. An open street map (OSM) from Berlin, PLZ 10969, Germany, is used. The intersection corresponds to Charlottenstrasse with Zimmerstrasse.
    The colour code indicates the path loss.}
    \label{fig:RaytracingExample}
    \vspace{-0.5cm}
\end{figure}
\begin{figure*}[t]
\centering
	\subfigure[Distance error from BS1.]{
		\includegraphics[trim={0.2cm 3cm 0.5cm 3cm},clip,width=158pt]{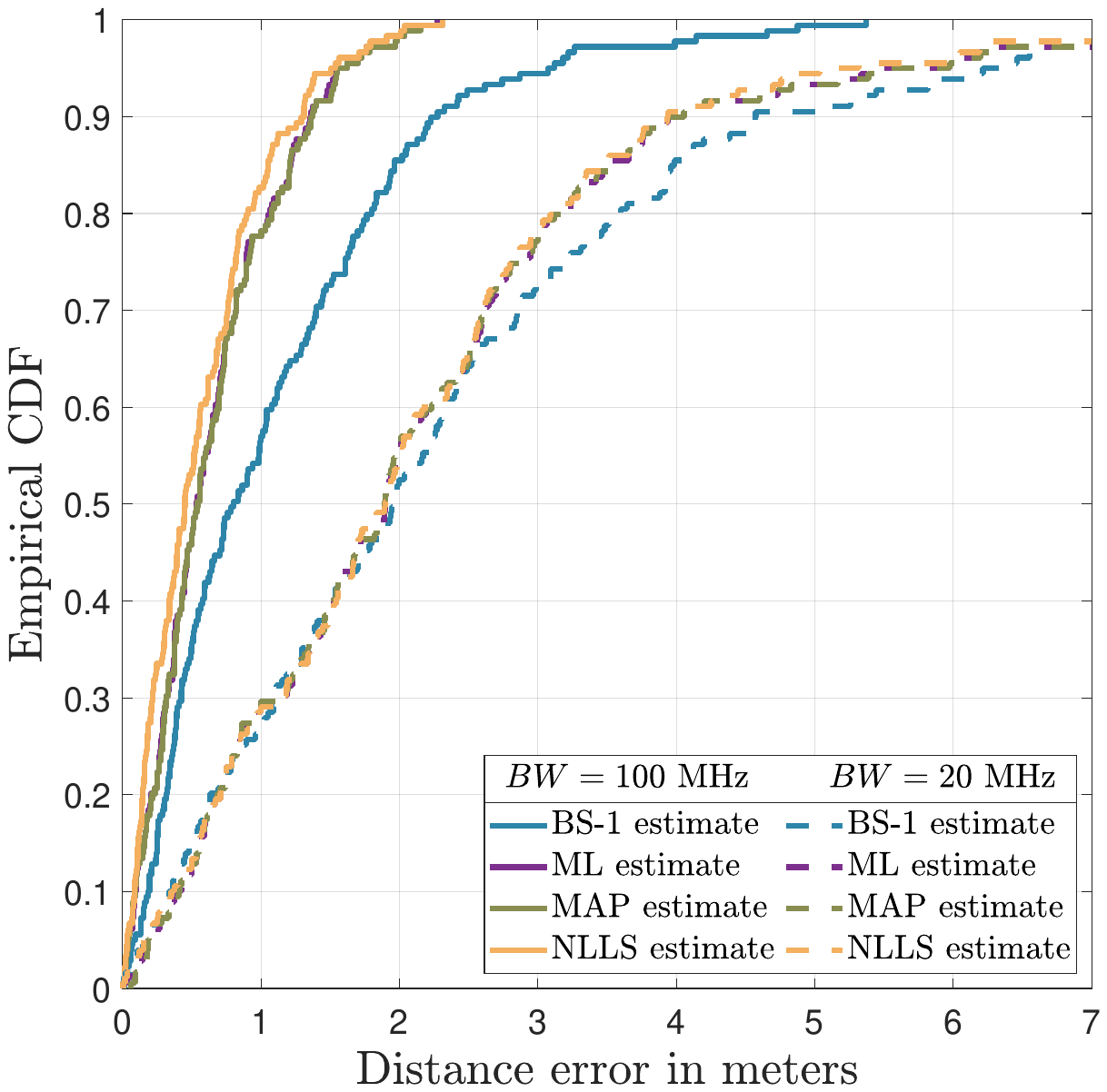}
        \label{subfig:DistanceErrorBS1K2}}
	\hfill
	\subfigure[Distance error from BS2.]{ 	
	\includegraphics[trim={0.2cm 3cm 0.5cm 3cm},clip,width=156pt]{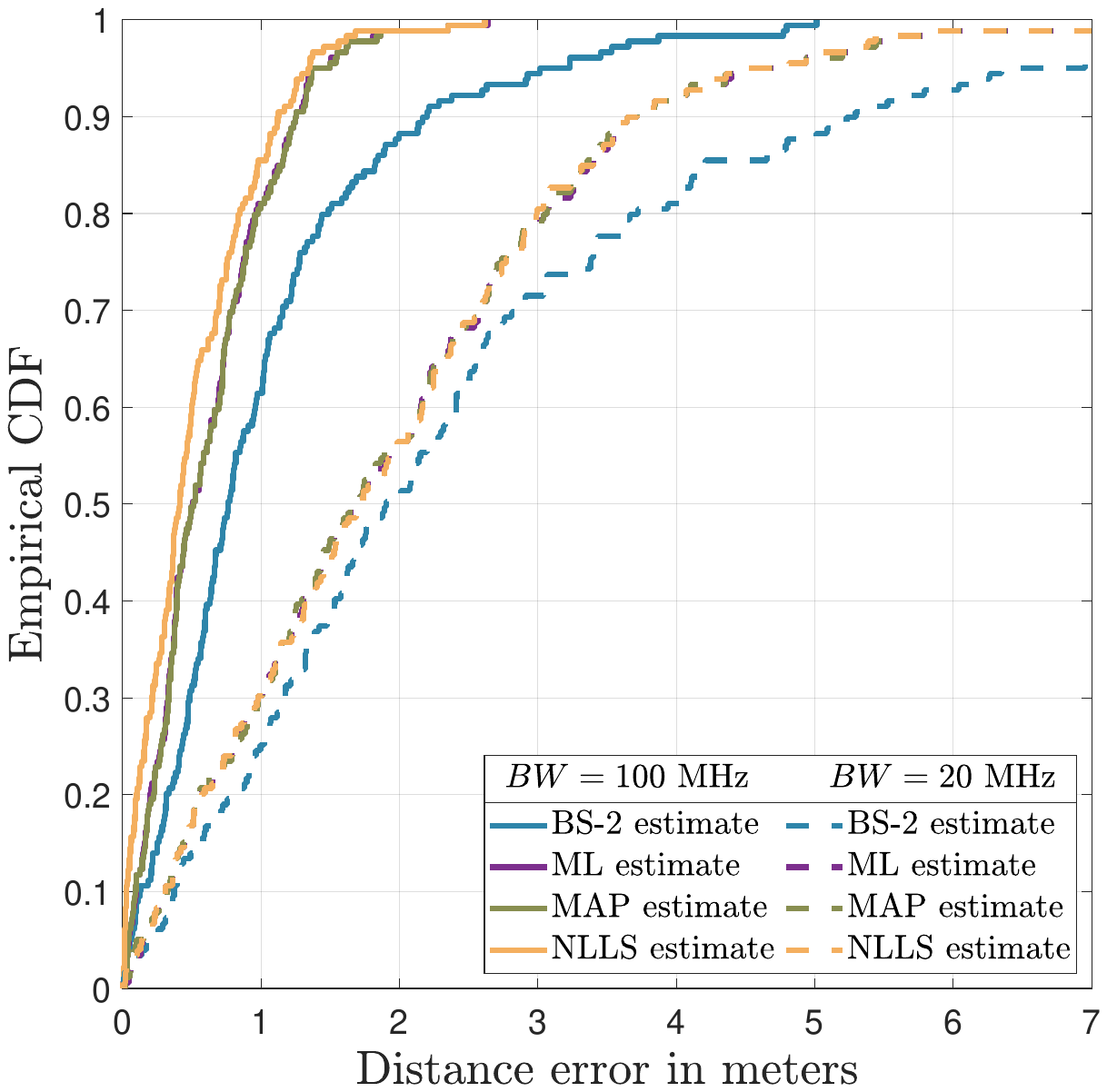}
    \label{subfig:DistanceErrorBS2K2}
	}
	\hfill
	\subfigure[Positioning error for $K = 2$.]{ 
	\includegraphics[trim={0.2cm 3cm 0.5cm 3cm},clip,width=156pt]{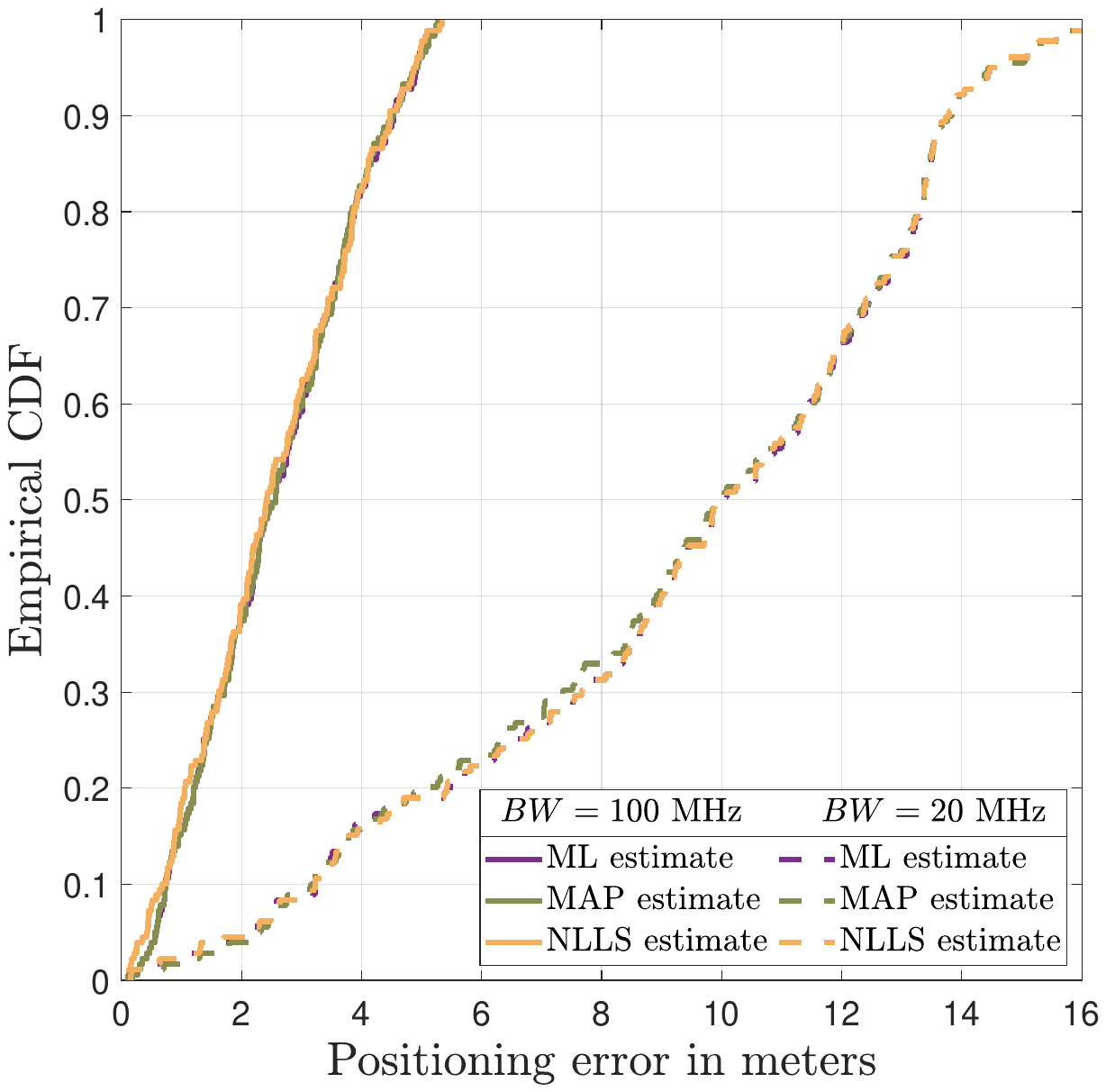}
    \label{subfig:PositioningErrorK2}
	}
	\caption{Distance and positioning error for different BW for $K = 2$.}
	\label{fig:Distance_K2}
    \vspace{-0.5cm}
\end{figure*}
\subsection{Fusion Algorithms}
We consider three fusion algorithms, namely, time-of-arrival (ToA) maximum likelihood (ML), ToA maximum a posteriori (MAP), and non-linear least square (NLLS) techniques, whose performances are analyzed in Section~\ref{sec:PerformanceEvaluation}. 

\subsubsection{ML Estimator} 
The ToA estimation error can be modelled as a zero-mean Gaussian random variable. The Log-likelihood (LL) function is given by  \cite{Henninger2022, Perez-Cruz2016}
\begin{equation}
    \text{LL}(\mathbf{x}) = \sum_{k\,\in\, \mathcal{K}} w_k \cdot \ln\Bigg(\frac{1}{\sqrt{2\pi}} \cdot \text{exp} \Big[- \frac{(\hat d_k - ||\mathbf{x}_k - \mathbf{x}||)^2}{2\,\sigma_k^2} \Big]\Bigg) \label{eq:LL}
\end{equation}
where $\sigma_k^2$ and $w_k$ are the variance of the Gaussian component and the weight of the $k$-th BS, respectively. Then, the location $\mathbf{x}$ is estimated by maximizing the LL function 
\begin{equation}
    (\hat{\mathbf{x}}_{\text{LL}}) = \arg\max_{\mathbf{x}}\,\,\text{LL}(\mathbf{x}) \label{eq:argmaxLL}
\end{equation}
Note that, in order to solve the equation above, the minimum number of BSs used needs to be equal to the number of unknown variables \cite{Perez-Cruz2016}.
\subsubsection{MAP Estimator}
The MAP estimator can be built by combining the ML estimator and a prior distribution.
Following \cite{Perez-Cruz2016}, the prior probability for the $k$-th BS can be modelled as a uniform distribution, defined as
\begin{equation}
    p_k(\mathbf{x}) = \frac{1}{||\mathbf{x} - \mathbf{x}_k|| + \epsilon} \label{eq:prior_k}
\end{equation}
where $\epsilon$ is a constant added for numerical stability. Furthermore, the prior of each $k$-th BS is independent of each other, i.e., $p(\mathbf{x}) = \prod_{k\,\in\, \mathcal{K}} p_k(\mathbf{x})$.
Subsequently, the Log-MAP function is given by
\begin{align}
    \text{LMAP}(\mathbf{x}) &= \sum_{k\,\in\, \mathcal{K}} w_k\Bigg[ \ln\Bigg(\frac{1}{||\mathbf{x} - \mathbf{x}_k|| + \epsilon} \Bigg) + \nonumber\\
    &\ln \Bigg(\frac{1}{\sqrt{2\pi}} \cdot \text{exp} \Big[- \frac{(\hat d_k - ||\mathbf{x}_k - \mathbf{x}||)^2}{2\,\sigma_k^2} \Big] \Bigg) \Bigg] \label{eq:LMAP}
\end{align}
and its estimator is given by
\begin{equation}
    (\hat{\mathbf{x}}_{MAP}) = \text{arg}\max_{\mathbf{x}}\, \text{LMAP}(\mathbf{x}) \label{eq:argmaxLMAP}
\end{equation}
The prior in the MAP estimator assigns less weight to targets that are further away, prioritizing closer targets. Setting equal weights leads to an unweighted LMAP estimator, where the prior serves as a weight.

\subsubsection{NLLS Estimator}
NLLS is a technique used when there is no information about the likelihood function and no feasible assumption \cite{Guvenc2009}. It is given as follows
\begin{equation}
    \hat{\mathbf{x}}_{\text{NLLS}} = \text{arg}\min_{\mathbf{x}} \sum_{k\,\in\, \mathcal{K}} w_k \, (\hat d_k - ||\mathbf{x}_k - \mathbf{x}||)^2 \label{eq:argminNLLS}
\end{equation}

It is important to notice that the weights $w_k,\,\forallK$, are normalized.

\subsection{Position Estimation} 

Before solving the equations \eqref{eq:argmaxLL}, \eqref{eq:argmaxLMAP}, and \eqref{eq:argminNLLS}, some parameters need to be estimated. The distance $\hat{d}_k$ is estimated using equations \eqref{eq:Distance_Velocity_Estimation_Periodogram} and \eqref{eq:PeakSelection}. 
It is imperative to understand that weights play a fundamental role in the fusion stage. As stated in \cite{Henninger2022}, obtaining the correct weights leads to less accuracy error. In addition, determining the weights implicates identifying and removing outliers so that the less reliable measurements are given less weight. In our approach, first, we do Gaussian fitting over the periodogram map of each BS (i.e. $ A_k(\cdot)$ in Eq. \eqref{eq:Periodogram}). The value of the variance of the fitted curve is used for $\sigma_k^2$, and the normalized amplitude value is used as weight ($w_k$), where the latter is proportional to the received power of LOS and most relevant NLOS paths. As the BS closer to the target will have a higher amplitude and smaller variance due to the dominating LOS path, the fusion algorithm gives more weight to the BSs closer to the target, minimizing the effect of multipath error in final position estimation.
Note that, in contrast to the approach in \cite{Henninger2022, Perez-Cruz2016}, where weights are calculated iteratively at each position, in our case, the weights are calculated in a simpler low-complex one-shot fashion. 

Since equations \eqref{eq:argmaxLL}, \eqref{eq:argmaxLMAP}, and \eqref{eq:argminNLLS} are non-convex, numerical optimization methods require a good starting point. Gradient-based algorithms have proven effective in solving this problem. Furthermore, in order to obtain a feasible and non-linear dependent solution, at least three BS estimates are needed to find the three unknown parameters of the target's position \cite{Guvenc2009, Henninger2022}.

\begin{table}[t]
    \small
	\centering
	\caption{System parameters}\label{tab:ScenariosandSetup}
	\renewcommand{\arraystretch}{1}
	{\begin{tabular}{|l|p{0.27\linewidth}|} \hline
			\textbf{Parameters} & \textbf{Values}\\ \hline 
            Carrier frequency, $f_c$ & $3.5$ GHz \\ \hline
            Transmit power, $P_{T}$ & $ 23 $ dBm \\ \hline
            Bandwidth, $BW$ & $20$ and $100$ MHz \\ \hline
            Subcarrier spacing, $\Delta f$ & $30$ kHz \\ \hline
			Number of OFDM symbols, $M$  & $ 500 $\\ \hline
			Number of BSs, $K$ & up to $3$\\ \hline
            Target RCS & $7$ dBsm \cite{Ge2023} \\ \hline
            Target Speed, $||\mathbf{v}||$ & $50$ km/h \\ \hline
	\end{tabular}}
	\renewcommand{\arraystretch}{1}
	\vspace{-.4cm}
\end{table}

\section{Performance Evaluation}\label{sec:PerformanceEvaluation}
\subsection{Setup and Scenarios}

To evaluate the localization performance of multi-monostatic sensing, we employ a realistic map model of Berlin at the Charlottenstrasse and Zimmerstrasse intersection. We simulate two scenarios. The first scenario considers two BSs at both ends of the street at an altitude of $10$ m, while the second scenario adds an extra BS located in the middle of the street at the rooftop, at $24$ m, as depicted in Fig.~\ref{fig:illustration_SystemModel}. 
All the BSs are deployed with omnidirectional antennas. We simulate a target moving throughout the street while the BSs estimate their relative distance to the target every $20$ ms. We use the Ray-Tracing for Wireless Communications, MATLAB \cite{MATLAB}, which considers all the multi-path components that are dependent on the geometry of the propagation environment and can be approximated to a realistic channel. 
This framework accounts for various electromagnetic phenomena, including reflections, refraction, diffraction, and diffuse scattering, resulting in a complex channel model.
Fig.~\ref{fig:RaytracingExample}, shows the simulation environment and the different paths of the signals sent by the BSs, for a given target's position. The BSs are shown in red and the target in blue.

For sensing purposes, an OFDM signal is sent in each interval, with 500 OFDM symbols and a sub-carrier spacing of $30$ kHz. The target is modelled as a car moving with a constant speed of $50$ km/h. and with an assumed height of $z = 1$ m. 
The rest of the system parameters are given in \mbox{Table \ref{tab:ScenariosandSetup}}. 

Note that solving equations \eqref{eq:argmaxLL}, \eqref{eq:argmaxLMAP}, and \eqref{eq:argminNLLS}, requires a comprehensive search over all potential target locations $\mathbf{x}$. However, since the target in our case is moving in a specific street, we restrict our search space to the particular street. This reduces the search space required to solve the optimization problem. Further, we consider at most three BSs, i.e., $|\mathcal{K}| = 3$. Consequently, we employ a heuristic approach to solve the equations.

\subsection{Simulation Results and Discussion}
We evaluate the effectiveness of the proposed fusion-based estimation framework for multi-monostatic passive object sensing in terms of two evaluation metrics: 
(i) distance error, which is the difference between the estimated distance after fusion $\hat{d}_{k, \,\iota}$ and the true distance $d^{\star}_k$ of the target to each BS, defined as $|\hat{d}_{k, \,\iota} - d^{\star}_k|$, where $\hat{d}_{k, \,\iota}$ is the distance from the estimated position to each BS expressed as $||\hat{\mathbf{x}}_{\iota} - \mathbf{x}_k||$, with $\iota \in \{\text{LL, LMAP, NLLS}\}$ and $k \in \mathcal{K}$,
and (ii) positioning error $||\hat{\mathbf{x}}_{\iota} - \mathbf{x}^{\star}||$, which is the difference between the estimated position after fusing the information from multiple BSs $\hat{\mathbf{x}}_{\iota}$ with $\iota \in \{\text{LL, LMAP, NLLS}\}$, 
as per Eq. \eqref{eq:argmaxLL}, \eqref{eq:argmaxLMAP}, and \eqref{eq:argminNLLS}, and the actual position of the target $\mathbf{x}^{\star}$. We evaluate the performance under the setup with $K = 2$ and $3$ BSs and bandwidth of $20$ and $100$ MHz. 

\subsubsection{Estimation error with fusion of two BSs}

\begin{figure}[t]
    \centering
    \includegraphics[trim={0.2cm 3cm 0.5cm 3cm},clip, width=0.7\linewidth]{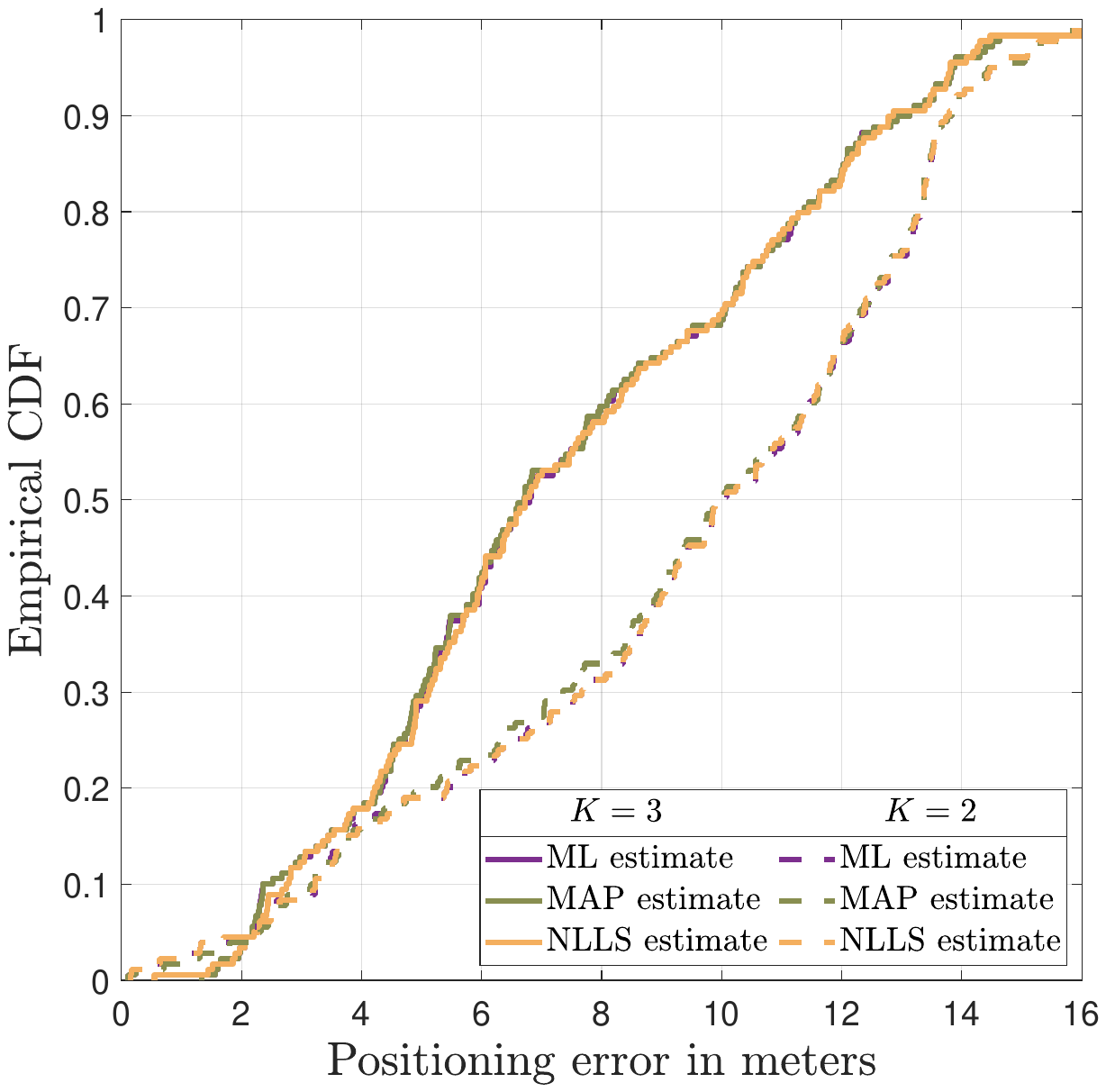}
    \caption{Positioning error for different number of BSs fused for \mbox{$BW = 20$ MHz.}}
    \label{fig:positioningError_20}
    \vspace{-0.25cm}
\end{figure}

\begin{figure}[t]
    \centering
    \includegraphics[trim={0.2cm 3cm 0.5cm 3cm},clip, width=0.7\linewidth]{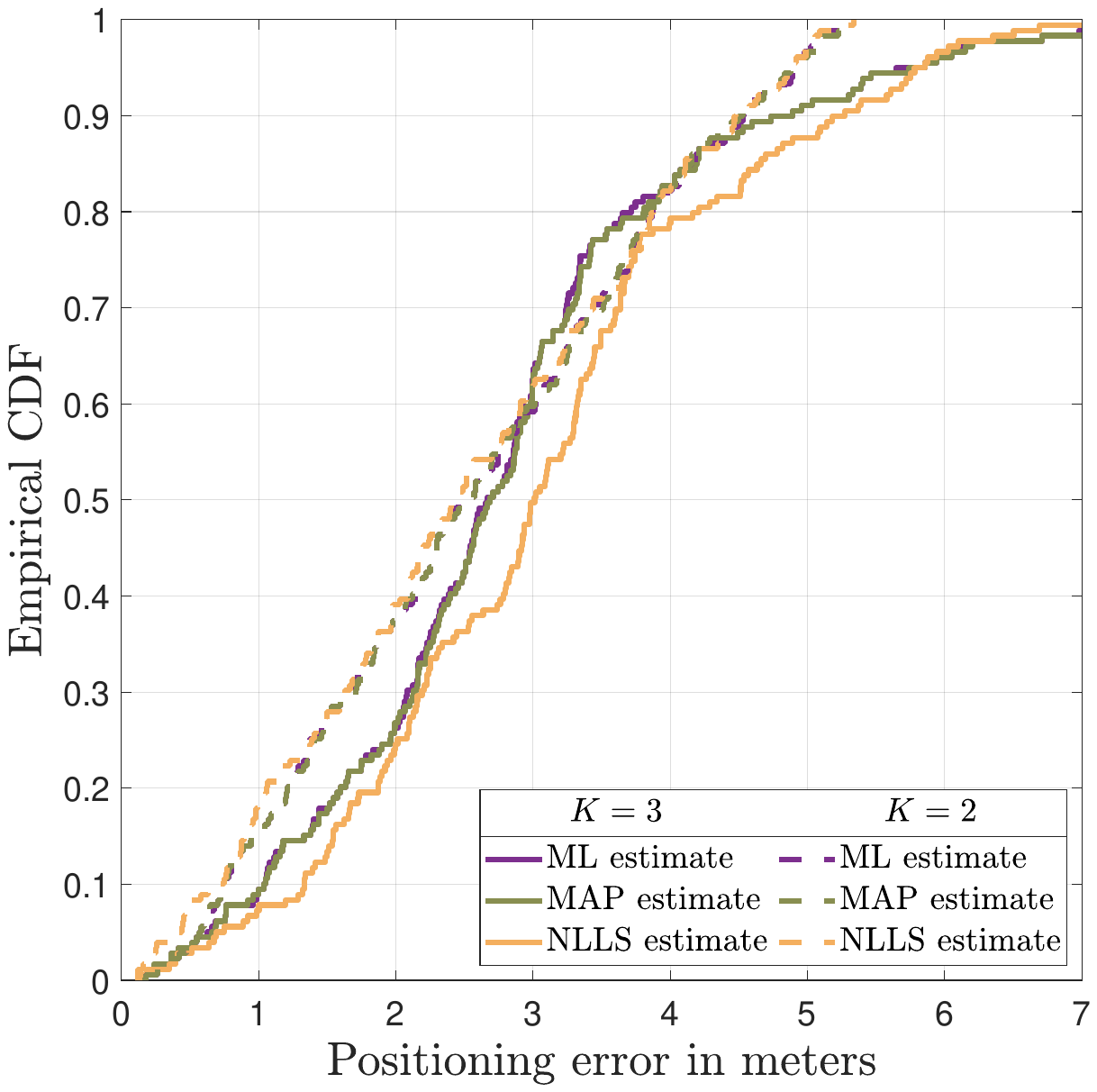}
    \caption{Positioning error for different number of BSs fused for \mbox{$BW = 100$ MHz.}}
    \label{fig:positioningError_100}
    \vspace{-0.5cm}
\end{figure}

Fig.~\ref{fig:Distance_K2} depicts the distance and positioning error when only BS1 and BS2 are fused (see Fig.~\ref{fig:RaytracingExample}).  Fig.~\ref{subfig:DistanceErrorBS1K2} and Fig.~\ref{subfig:DistanceErrorBS2K2} illustrate the empirical cumulative density function (CDF) of the distance error, where the different fusion techniques are compared with the estimate of BS1 and BS2 respectively under different bandwidths. It can be clearly observed in Fig.~\ref{subfig:DistanceErrorBS1K2} that there is a substantial gain in the BS1 estimates for a bandwidth of 100 MHz. This can be understood from Eq.~\eqref{eq:Distance_Velocity_Estimation_Periodogram}, where the resolution of the estimation is proportional to the bandwidth given for sensing.
When $BW = 20$ MHz, we observe a slight gain when using fusion compared to the single BS1 estimate, and there is no considerable difference between the fusion algorithms (ML, MAP, and NLLS). 
In the case of $BW = 100$ MHz,  there is a considerable gain between the single BS1 estimate and the fusion techniques. 
From all three fusion approaches, NLLS performs slightly better, whereas ML and MAP have similar performances, stating that the prior used is diffuse and gives no additional information. 
One reason for the better performance of NLLS is that the distance estimation error no longer follows a zero-mean Gaussian distribution due to the multipath.
It is worth noting that for $90\%$ of confidence, there is a gain in terms of range accuracy of almost \mbox{$1$ m} between the BS estimate and the fusion algorithms. The results confirm that cooperative fusion of the estimates of two BSs can substantially reduce the estimation error.
This trend is similar for BS2 in Fig.~\ref{subfig:DistanceErrorBS2K2}. Compared to Fig.~\ref{subfig:DistanceErrorBS1K2} we observe that for \mbox{$BW=20$ MHz}, the gain in range accuracy of the fusion-based estimations compared to the BS2 estimation is more prominent. This is because BS1 receives fewer dominating NLOS components of echoes than BS2 due to its location. This demonstrates that the fusion-based algorithm improves the range estimation accuracy and can compensate the effect of a BS being at a "unfavourable" multipath-rich location.
Fig.~\ref{subfig:PositioningErrorK2} depicts the empirical CDF of the positioning error for different bandwidths. Here, we clearly observe the effect on the available bandwidth on the positioning error. For $90\%$ of confidence, there is a gain of around $10$ m when increasing the BW from $20$ MHz to $100$ MHz.
Even though Figs \ref{subfig:DistanceErrorBS1K2} and \ref{subfig:DistanceErrorBS2K2} NLLS curves depict a better performance, the three fusion techniques have almost the same performance in terms of positioning error. 
It is important to note that as shown in Fig.~\ref{fig:Distance_K2}, ML and MAP have similar performance. This is due to the uniform distribution of the prior function considered by the MAP estimator, which does not account for the impact of bias on the estimation of the BS position. For different distributions, such as exponential and Gaussian, it may result in a higher gain than ML estimation.
\subsubsection{Estimation error with fusion of three BSs}
Figs.~\ref{fig:positioningError_20} and \ref{fig:positioningError_100}  show the effect of adding one extra BSs in the middle of the street, shown as BS3 in Fig.~\ref{fig:RaytracingExample}, on the estimation error for $BW = 20$ MHz and \mbox{$BW = 100$ MHz}, respectively. 
In Fig.~\ref{fig:positioningError_20}, there is a considerable gain when three BSs are fused in comparison to the case when two BSs are fused, mainly due to the low resolution owning to the small bandwidth of \mbox{$20$ MHz}. However, when increasing the bandwidth to \mbox{$100$ MHz}, as depicted in Fig.~\ref{fig:positioningError_100}, the performance is similar to the case when there are only two BSs.
The lack of gain is mainly due to BS3 being located in the middle of a rooftop and suffering from multipath more prominently than the other two BSs. This underlines that the quality of the estimation depends on both the location of each BS and the method employed for the selection of the BS for fusion.  


\section{Conclusions}\label{sec:Conclusions}
This paper proposes a two-stage framework for passive target positioning in the FR1 band using a 5G NR-based multi-monostatic sensing approach. We investigated the distance and positioning error of the proposed framework under different BW allocations for three distinct fusion methods. Our results indicate that increasing the bandwidth at the BSs results in a significant improvement in distance and positioning accuracy. Further, we show that, under low resolution, the fusion of BSs' estimates can result in higher accuracy. However, under higher resolution and depending on the location of the BSs, this gain is compromised based on the multipath condition of each BS. 
Therefore, the location and multipath condition of each BS significantly affect the fusion's efficacy.
In our future work, we plan to include multi-target localization and investigate BS selection mechanisms in severe multipath environments using sophisticated fusion methods.


\small
\section*{Acknowledgment}
This work was partially funded by the Federal Ministry of Education and Research Germany within the project ”KOMSENS-6G” under grant 16KISK128.
\bibliographystyle{IEEEtran}
\bibliography{icc_multimono.bib}

\end{document}